\documentstyle[prl,aps]{revtex}
\begin{document}

\vskip 4mm

\centerline{\large \bf Formation of Clusters of Interstitial Carbon Atoms}
\centerline{\large \bf in Graphite due to Deformation Interaction}
\centerline{\large \bf and their Spatial Arrangement}

\vskip 2mm

\centerline{V.F.Elesin and L.A.Openov}

\vskip 2mm

\centerline{\it Moscow State Engineering Physics Institute
(Technical University),}
\centerline{\it Moscow, 115409, Russia}

\vskip 4mm

\begin{quotation}

Formation of clusters of interlayer interstitial carbon atoms in graphite is
studied by means of molecular dynamics simulation. It is shown that the
deformation potential is attractive for interstitials located in one
interlayer region and repulsive for interstitials located in different
interlayer regions. As a result, relatively small interstitial clusters are
formed which are arranged in a checkerboard-like order.

\end{quotation}

\vskip 6mm

{\bf 1. Introduction}

\vskip 2mm

Irradiation of solids by neutrons, ions, electrons and other particles
results in the formation of irradiation-induced defects. The processes of
defects creation are accompanied by the processes of defects annihilation,
mainly due to recombination of vacancies and interstitials. Hence, one can
expect that a stationary defect state should be reached, such a state being
characterized by the relatively low defect concentration $n_d$. However, the
expected saturation of $n_d$ is not seen in the experiment
\cite{Iwata,Kelly}. For example, the volume of graphite increases
monotonously under irradiation up to very high doses. Hence, there exists a
mechanism that prevents the recombination.

One possible explanation of recombination suppression is the formation of
clusters of point defects at the sacrifice of deformation interaction
\cite{Elesin1,Elesin2}. As is known (see, e.g., \cite{Kosevich}),
interstitials stretch the lattice, while vacancies compress the lattice. The
stretched domains are attractive for interstitials and repulsive for
vacancies. As a result, clusters of interstitials and clusters of vacancies
should be formed. The formation of such clusters has been predicted in
\cite{Elesin1,Elesin2} by the example of isotropic crystals. Spatial
separation of interstitial clusters and vacancy clusters suppresses the
recombination. In principle, this allows one to explain the accumulation of
defects (and, hence, structural changes and swelling) up to high radiation
doses.

We note that the region of deformation interaction in isotropic crystals is
rather small (of the order of one atomic volume), i.e., such an interaction
is point-like in the continuum limit. The range of deformation interaction in
anisotropic crystals (e.g., in layered graphite) may be much longer than in
isotropic ones. Indeed, interstitials placed between the graphite layers
cause the strong deformation of those layers \cite{Coulson,Taji}. Hence, the
formation of interstitial clusters may be expected to be more pronounced in
anisotropic crystals.

Recently a number of experimental results has been obtained that lend support
to the physical picture described above. In particular, it was shown
experimentally and theoretically \cite{Marton1} that krypton atoms in
graphite form clusters located between the layers. Those atoms come close
together due to forces acting on them from the deformed layers. According to
numerical simulations of 1, 2, and 3 krypton atoms placed into the interlayer
region, the deformation forces are so strong that the distance between the
krypton atoms in clusters appears to be shorter than that in the bulk krypton
crystal \cite{Marton1}.

The purpose of this work is the theoretical study of deformation interaction
of interstitial carbon atoms located between the graphite layers, as well as
the study of formation of interstitial clusters between several layers. As a
first step, we make use of a simple model with empirical interatomic
potential. By means of molecular dynamics simulations, it is shown that
interstitials located between the nearest layers attract to each other. On
the other hand, interstitials located between the different pairs of graphite
layers (i.e., in different interlayer regions) repel each other. This
repulsion sets limits on the size of interstitial clusters and prevents their
transformation into "superplanes", contrary to common opinion. The relatively
small interstitial clusters are arranged in a checkerboard-like order.

\vskip 6mm

{\bf 2. Deformation Interaction of Two Interstitial Carbon Atoms in Graphite}

\vskip 2mm

Numerical simulation was carried out by molecular dynamics technique making
use of empirical interatomic potentials for interaction between the carbon
atoms of graphite layers as well as between the interstitial carbon atoms and
the atoms of graphite layers \cite{Taji}. Parameters of those potentials were
chosen by fitting the calculated values of interatomic distances and elastic
moduli to their experimental values in defect-free graphite. Besides, the
calculated value of self-energy of a single interstitial, $E_{is}=$ 1.8 eV,
agrees well with experimental data and theoretical estimates
\cite{Coulson,Thrower}. The covalent interaction between the interstitials
was taken into account approximately as hard-core interaction of solid
spheres having the radius $d/2$, where $d=$ 1.42 {\AA} is the covalent bond
length within graphite layers.

The simulation was carried out for the 1320-atom crystallite consisted of
6 layers, each of which included 220 carbon atoms. In order to keep the
stability of the crystallite, we fixed the atoms of upper and lower layers,
as well as the atoms on the peripheries of layers. All other atoms were
allowed to relax after the interstitials have been inserted into the
crystallite.

Fig.1 shows the results of calculation of atomic configurations (after
relaxation) for the case that two interstitials are located between the
nearest graphite layers (i.e., in one interlayer region) at a fixed in-plane
distance $r$ from each other. The energy $U$ of deformation interaction was
defined as follows:

\begin{equation}
U(r) = E(N+2,r) - E(N+1,{\bf r}_1) - E(N+1,{\bf r}_2) + E(N),
\label{U(r)}
\end{equation}
where $E(N)$ is the total energy of $N$-atom defect-free crystallite,
$E(N+1,{\bf r}_1)$ and $E(N+1,{\bf r}_2)$ are the total energies (after
relaxation) of the crystallite containing an interstitial in the position
${\bf r}_1$ and ${\bf r}_2$ respectively, $E(N+2,r)$ is the total energy
(after relaxation) of the crystallite containing two interstitials in the
positions ${\bf r}_1$  and ${\bf r}_2$, and $r$ is the in-plane distance
between the interstitials.

The dependence of the deformation energy $U$ on $r$ is shown in Fig.2. One
can see that $U$ is negative at $r<8$ {\AA}, decreases monotonously as $r$
decreases, and reaches the value $\approx$ -1 eV at $r=d$ (a small maximum
$U\approx$ 0.03 eV at $r\approx$ 10 {\AA} may be due to finite size effects).
So, in accordance with the mechanism considered in \cite{Elesin1}, we see
that a rather strong mutual attraction of interlayer interstitials does exist
since it is energetically favorable for interstitials to be located in the
domain of stretched lattice. Besides, the range of deformation interaction is
rather long ($\approx$ 10 {\AA}), as would be expected for anisotropic
crystals.

Fig.3 shows the results of calculation of atomic configurations (after
relaxation) for the case that two interstitials are located between the
different pairs of graphite layers (i.e., in different interlayer regions) at
a fixed in-plane distance $r$ from each other. The dependence of the
deformation energy $U$ on $r$ is shown in Fig.4. One can see that $U$ is
positive at $r<6$ {\AA}. This corresponds to the mutual repulsion of two
interstitials. Note that $U$ increases monotonously as $r$ decreases, and
reaches the value $\approx 3$ eV at $r=0$ (a small minimum
$U\approx$ -0.05 eV at $r\approx 8$ {\AA} may be due to finite size effects).

\vskip 6mm

{\bf 3. Formation of Clusters of Interstitial Carbon Atoms in One Interlayer
Region}

\vskip 2mm

Since two interstitials located between the nearest graphite layers attract
to each other at the sacrifice of deformation interaction, one can expect
that several interstitials located in one interlayer region should come
closer together and form a cluster. In order to study the dynamics of
interstitial clusters formation, we performed the calculations as follows.
Initially (at $t$ = 0) several interstitials were randomly distributed
between two layers of the crystallite. Then all atoms were allowed to relax,
including the interstitials.

Fig.5 shows the dynamics of 7-interstitial cluster formation. At the first
stage ($t=200$) three clusters composed of 2, 2, and 3 interstitials are
formed. At the next stage ($t=800$) two of those clusters fuse into a
5-interstitial cluster. Finally, the 7-interstitial cluster is formed. Thus,
all interstitials merge together into a single cluster due to deformation
interaction. From Fig.5 one can see that a characteristic time of 2- and
3-interstitial clusters formation is an order of magnitude shorter than the
time of fusion of those clusters into a larger one. This fact indicates that
the mobility of a cluster decreases drastically with the number of
interstitials in it.

We note that the 7-interstitial cluster has a trigonal structure, as opposed
to a hexagonal structure of the graphite layer. The reason is that the
interaction between the interstitials was taken into account in a rather
crude way.

Similar results were obtained for several other initial distributions of
interstitials in the interlayer region, as well as for different numbers of
interstitials in the region. It turns out that the resulting shape of an
interstitial cluster depends on the initial distribution of interstitials,
while the structure of the cluster is always trigonal. It can be shown that
the covalent bonding between the interstitials results in formation of a
fragment of graphite layer having a hexagonal structure.

\vskip 6mm

{\bf 4. Formation of Clusters of Interstitial Carbon Atoms in Three
Interlayer Regions}

\vskip 2mm

As shown above, interstitials located in one interlayer region form a large
cluster. The reason is that the deformation potential is attractive for such
interstitials. As the number of interstitials in the cluster increases, the
cluster transforms into a new graphite layer.

However, once interstitials are created throughout the whole sample, the
deformation repulsion of interstitials located in different interlayer
regions should have a profound effect on the process of interstitial clusters
formation. In order to study this effect, we performed the calculations for
the case that initially (at $t=0$) as many as 39 interstitials were randomly
distributed over three interlayer regions of the crystallite (with the
constraint of 13 interstitials in each interlayer region).

Fig.6 shows the dynamics of interstitial clusters formation. We stress that
in none of three interlayer regions do all 13 interstitials merge into a
single cluster, contrary to the case of cluster formation in one interlayer
region (see Fig.5). It is seen from Fig.6 that two clusters have been formed
in the middle and upper regions each, while four clusters have been formed in
the lower region. We stress that clusters are arranged in a checkerboard-like
order. The reason is the deformation repulsion of interstitials located in
different interlayer regions.

\vskip 6mm

{\bf 5. Conclusions}

\vskip 2mm

1. We have shown that there is strong deformation interaction of interstitial
carbon atoms in graphite. This interaction is attractive for interstitials
located in one interlayer region and repulsive for interstitials located in
different interlayer regions. A characteristic value of the deformation
energy $U$ is about 1 eV. The dependence of $U$ on the in-plane distance
between two interstitials was calculated.

2. The deformation attraction of interstitial carbon atoms was demonstrated
to result in formation of interstitial clusters in interlayer regions.

3. It was shown that deformation repulsion of interstitials prevents the
transformation of interstitial clusters into new graphite layers because of
formation of relatively small clusters arranged in a checkerboard-like order.
We believe this effect to be the reason for recombination suppression in
graphite because of spatial separation of interstitial clusters and vacancy
clusters. This is a possible explanation for the absence of saturation of
defect concentration in graphite up to high radiation doses.

Finally, it is worth noting that recently several researches
\cite{Porte1,Coratger,Porte2,Marton2,Matsukawa} have reported STM images of
defect structures on the surfaces of ion-irradiated highly-oriented
pyrolythic graphite (HOPG). Analysis of published experimental results points
to an important role of processes of interstitial clusters formation due to
deformation interaction.

\vskip 6mm

{\bf Acknowledgments}

The work was supported by the International Science and Technology Center,
Project No 467.

\newpage
\centerline{\bf FIGURE CAPTIONS}
\vskip 2mm

Fig.1. Distortion of graphite lattice by two interstitials located between
the nearest graphite layers at the in-plane distance (A) $r=$ 9.2 {\AA} and
(B) $r=$ 1.7 {\AA} from each other.

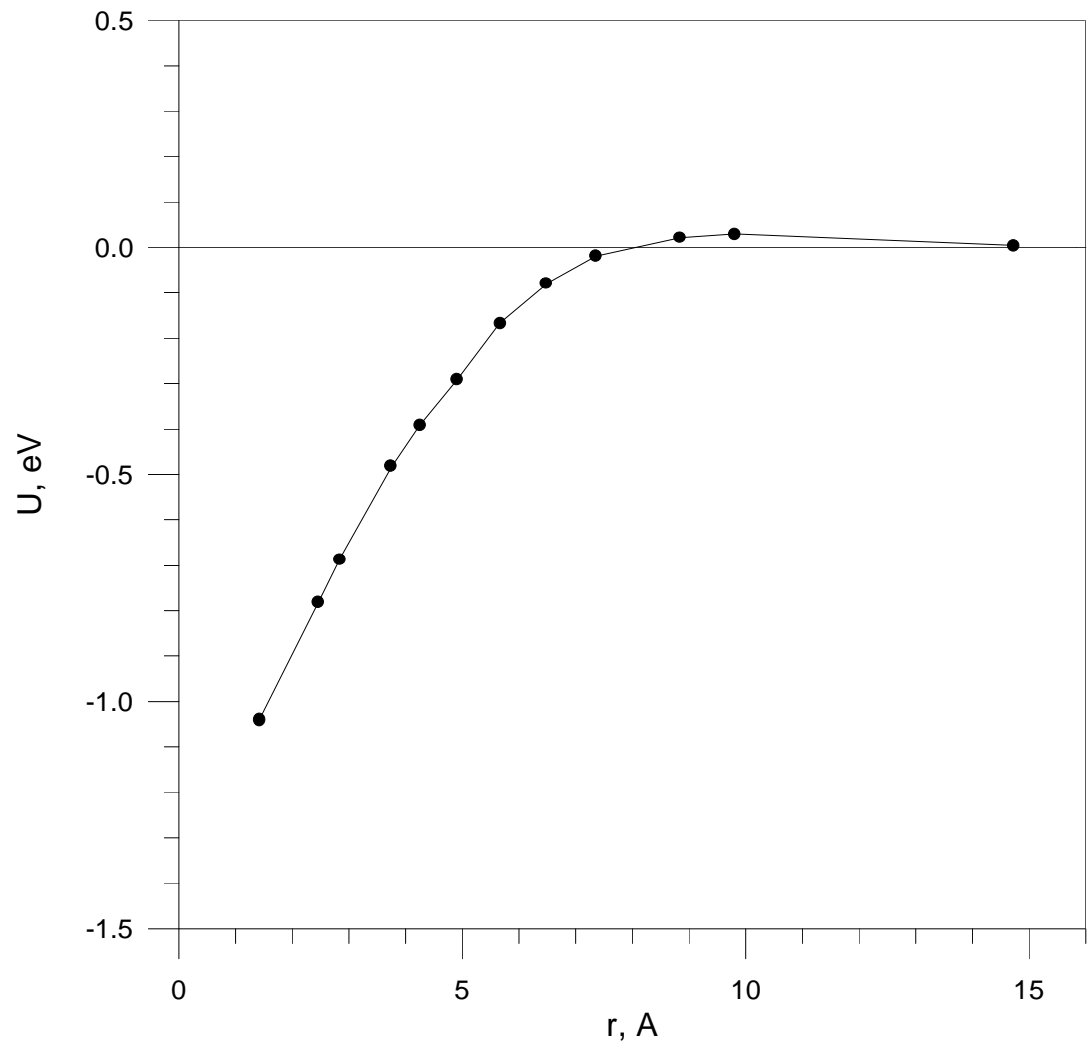
Fig.2. Dependence of the deformation energy $U$ on the in-plane distance $r$
between two interstitials located between the nearest graphite layers, see
Fig.1. The dots are the results of calculation. The line is a guide to the
eye.

Fig.3. Distortion of graphite lattice by two interstitials located between
the different pairs of graphite layers at the in-plane distance
(A) $r=$ 9.8 {\AA} and (B) $r=0$ from each other.

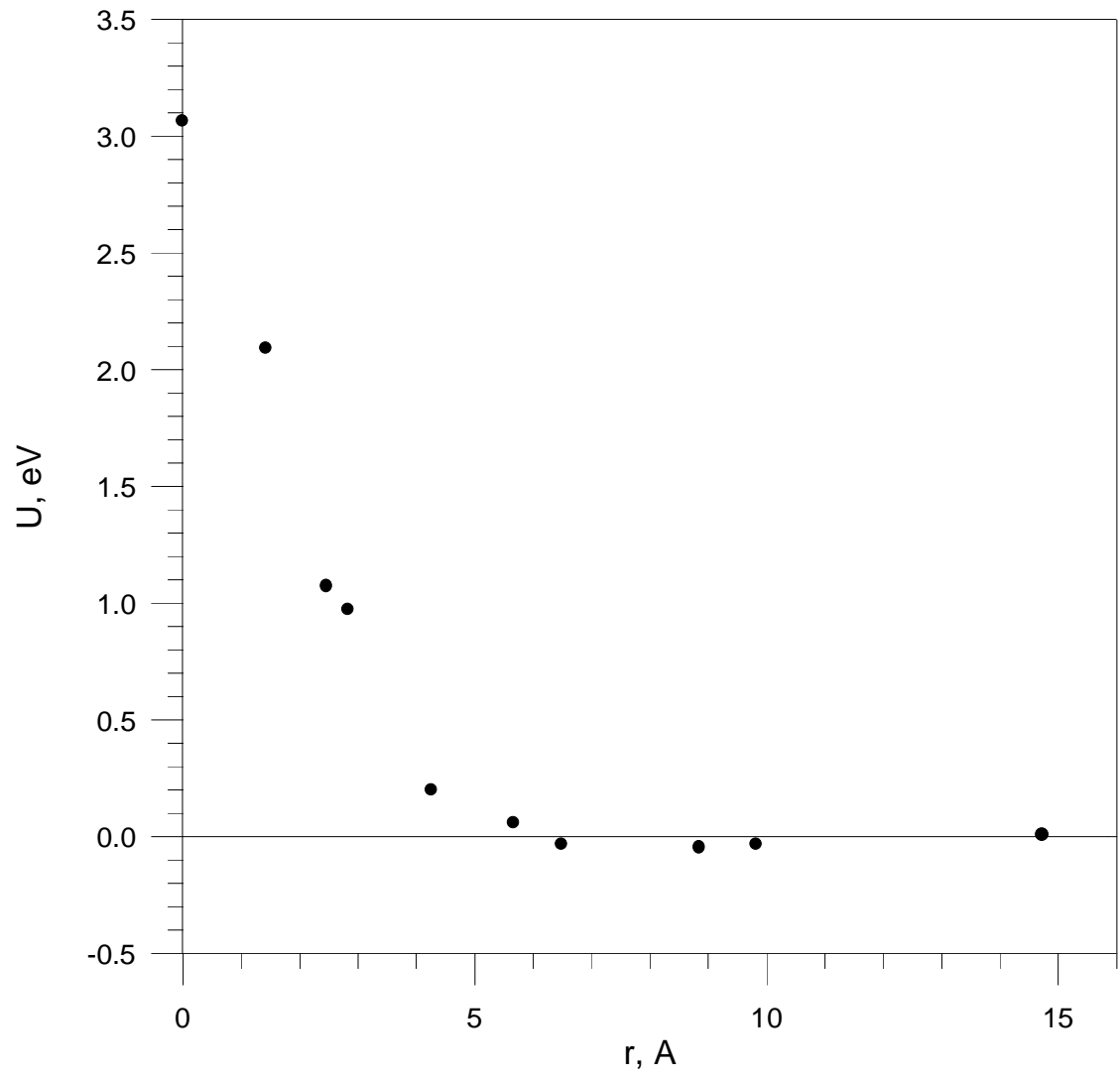
Fig.4. Dependence of the deformation energy $U$ on the in-plane distance $r$
between two interstitials located between the different pairs of graphite
layers, see Fig.3.

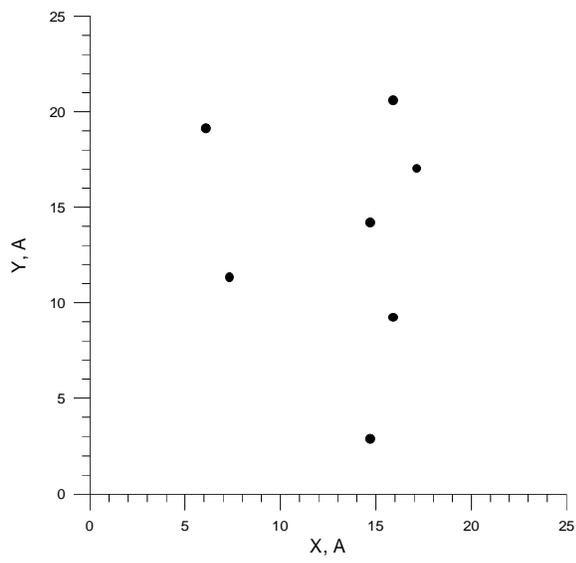
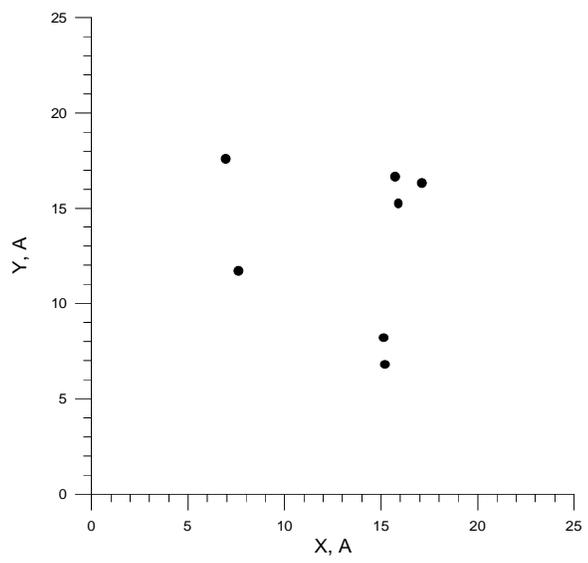
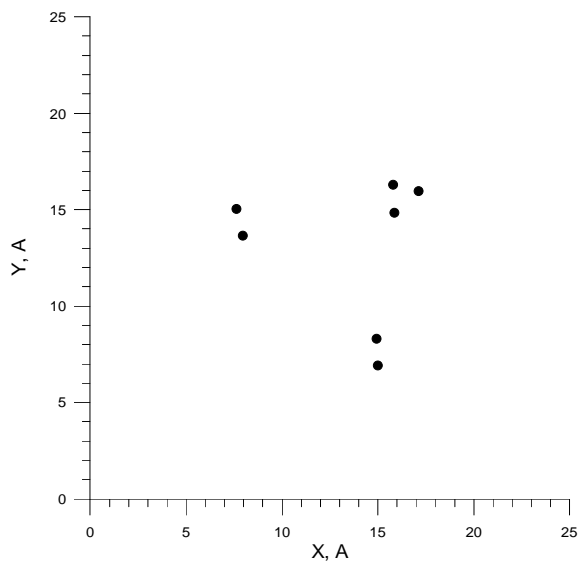
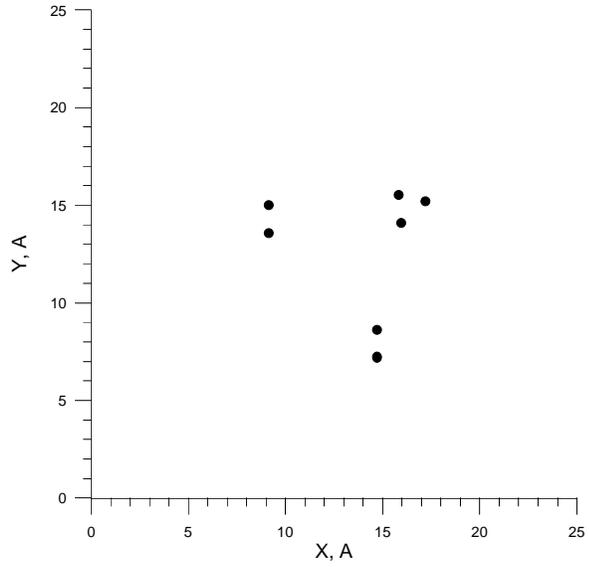
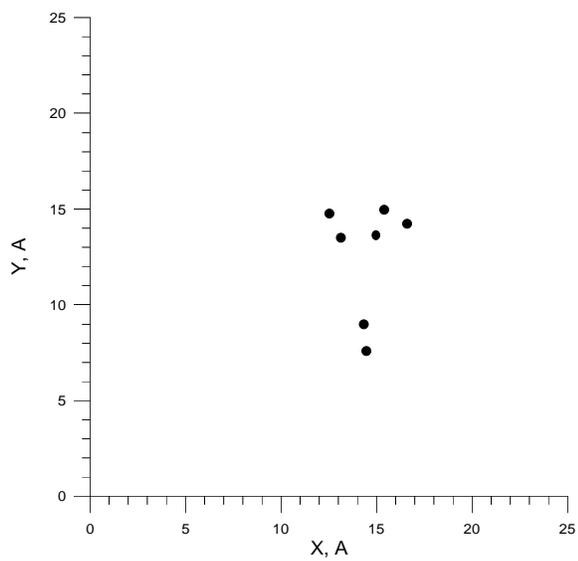
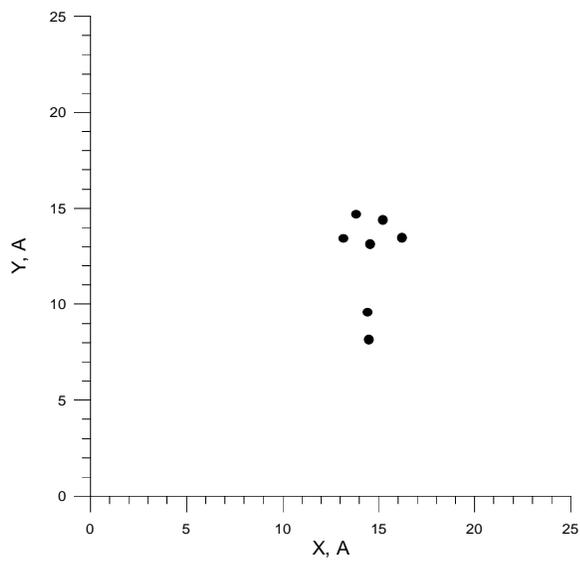
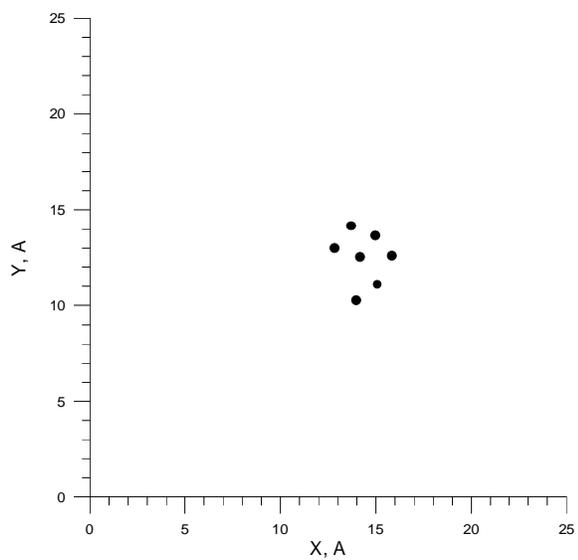
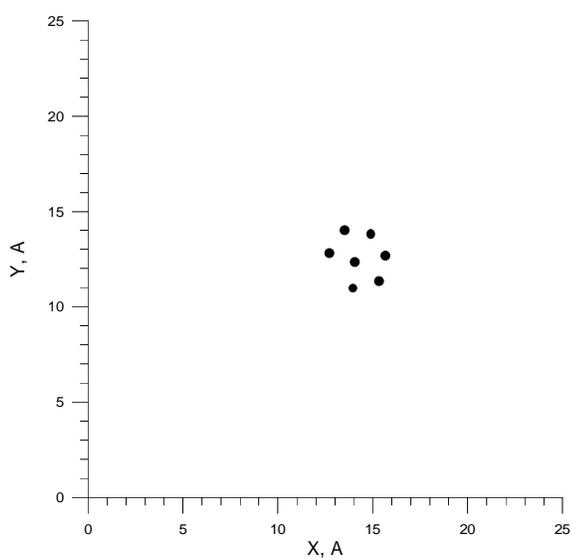
Fig.5. Dynamics of interstitial cluster formation in one interlayer region.
Top view. Atoms of graphite layers are not shown. Time $t$ is in arbitrary
units.

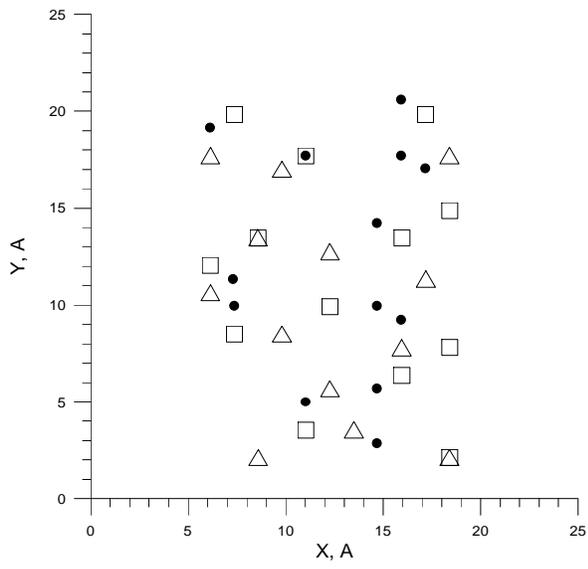
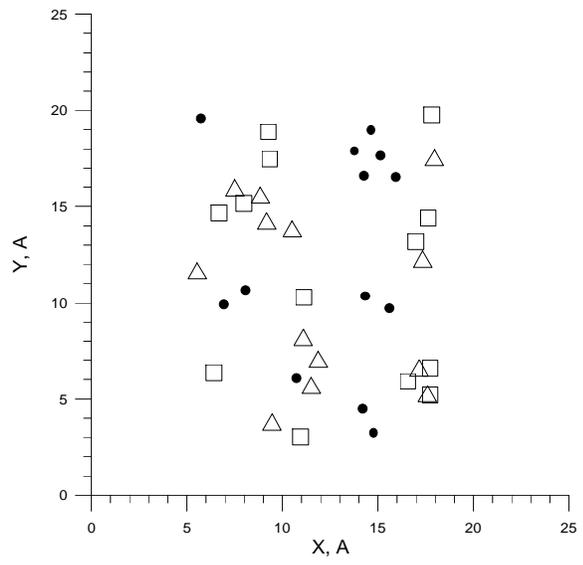
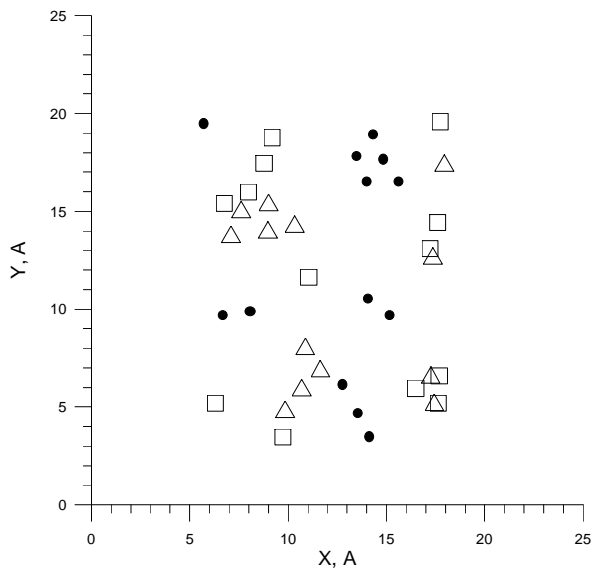
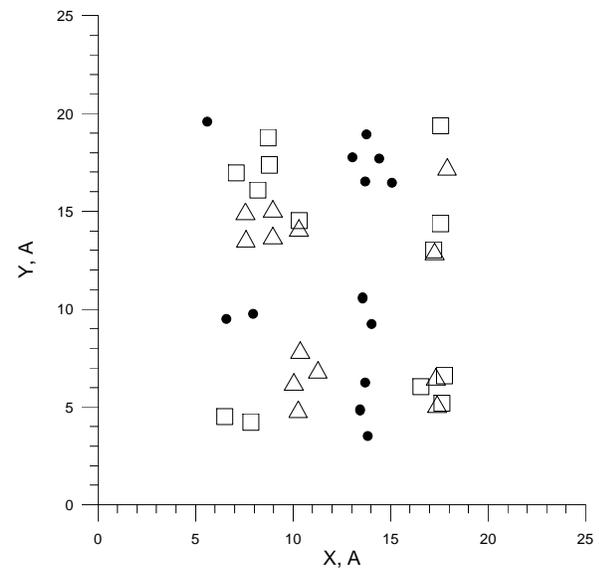
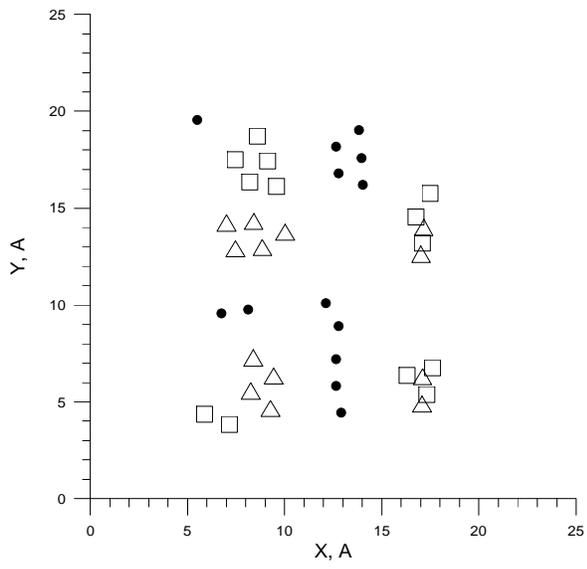
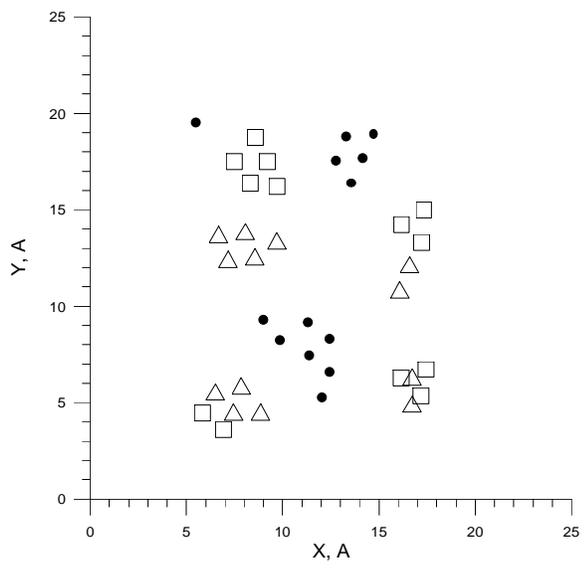
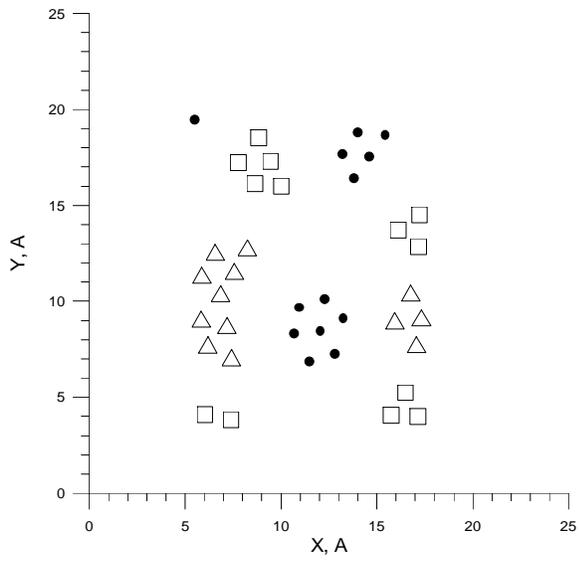
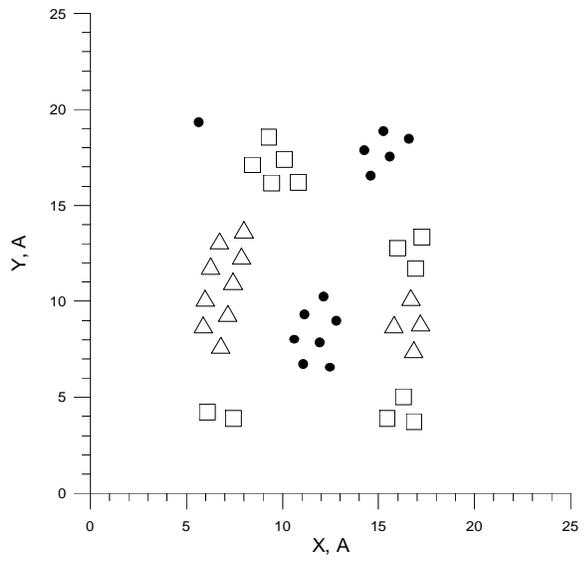
Fig.6. Dynamics of interstitial clusters formation in three interlayer
regions. Top view. Squares, circles, and triangles show the interstitials in
lower, middle, and upper regions respectively. Atoms of graphite layers are
not shown. Time $t$ is in arbitrary units.

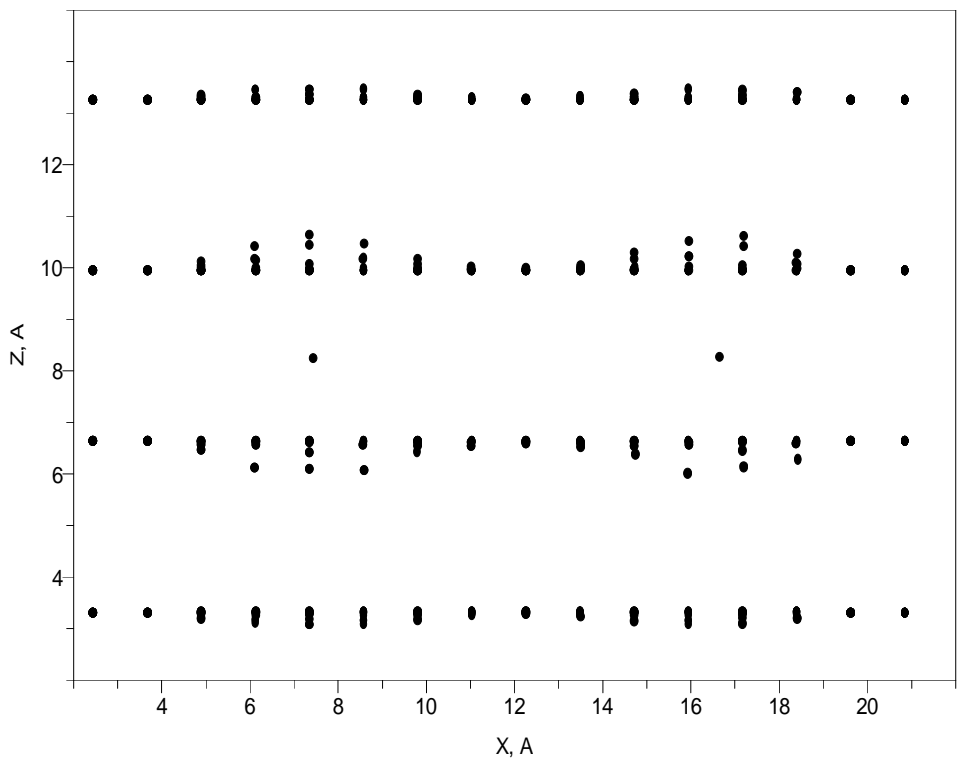
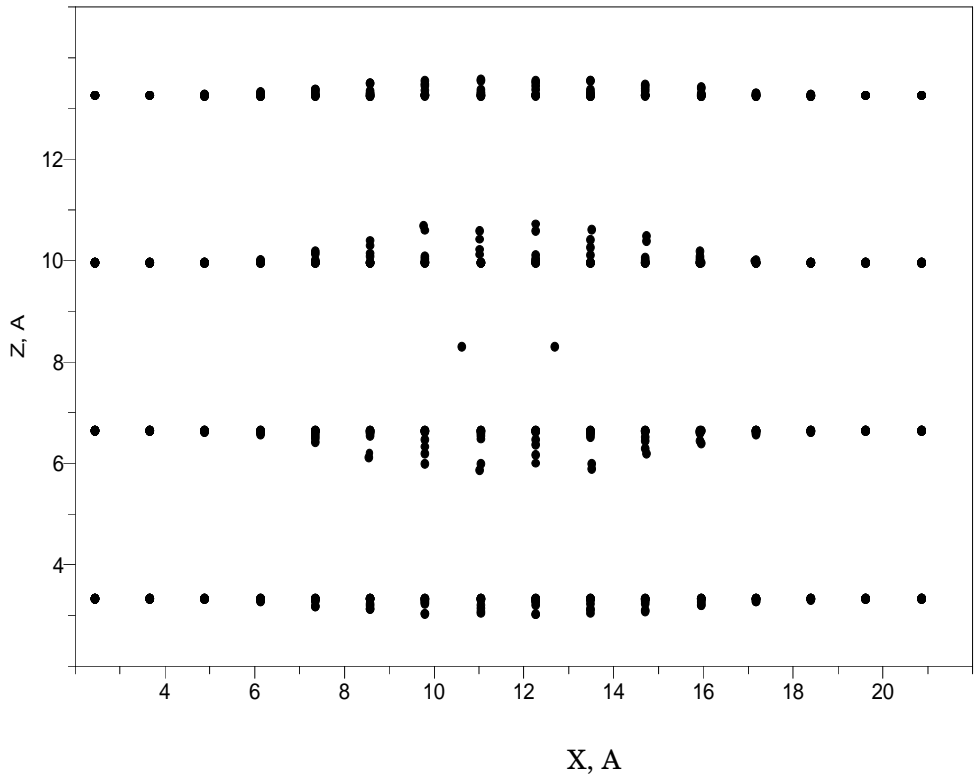
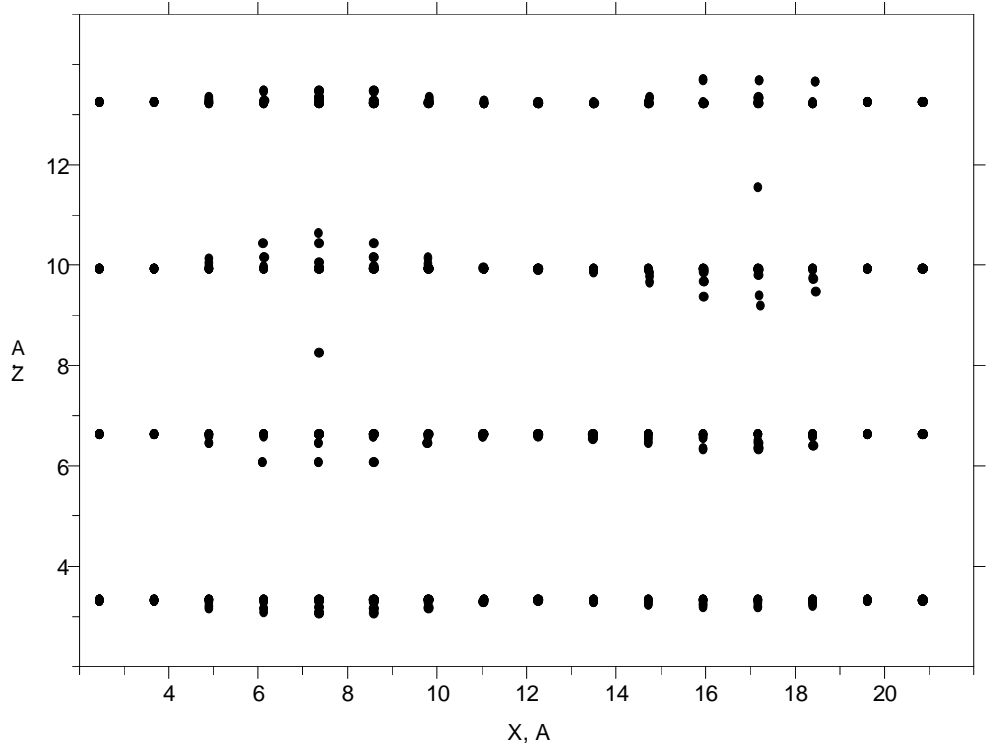
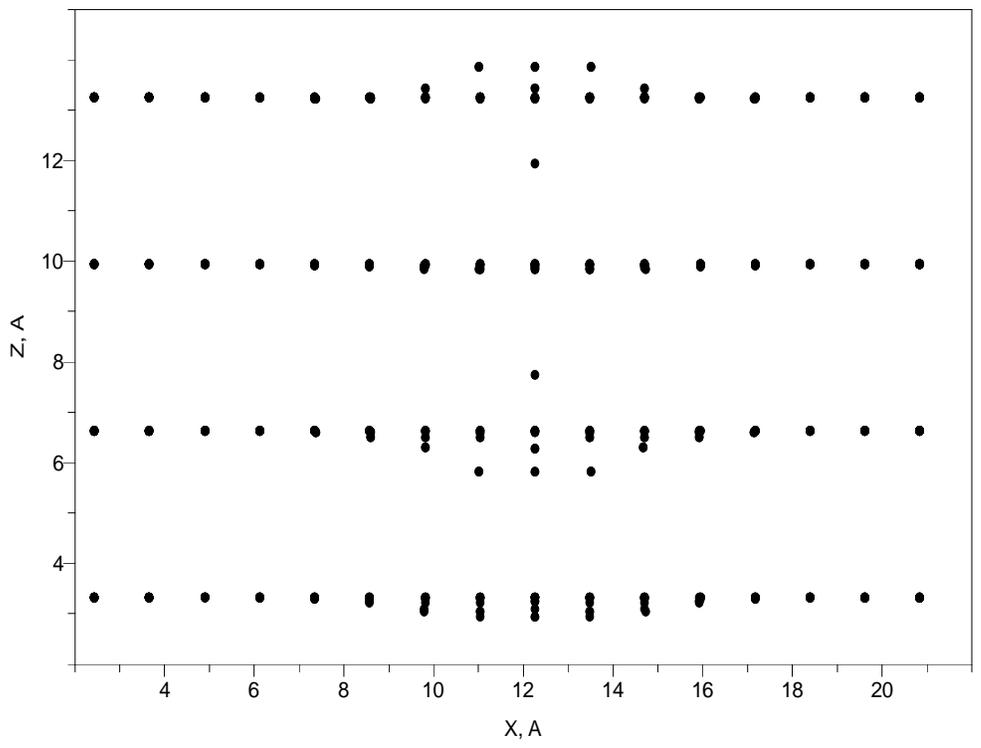
\end{document}